\documentclass[aps,pra,showpacs,twocolumn,superscriptaddress]{revtex4-1}

\usepackage{amsfonts,amsmath,amssymb,graphicx}
\usepackage[usenames,dvipsnames]{xcolor}
\usepackage{pdfpages}
\usepackage{footmisc}

\begin{document}

\title{Optimal state transfer of a single dissipative two-level system}
%

\author{H. Jirari}
\email{hjirari@hotmail.com}
\affiliation{13, rue Saint Fulbert, 69008 Lyon, France}
\author{N. Wu}
\affiliation{Department of Chemistry, Princeton
         University, Princeton, New Jersey 08544, USA}
%


\begin{abstract}
Optimal state transfer of a single
two-level system (TLS) coupled to an Ohmic boson bath via off-diagonal
TLS-bath coupling is studied by using optimal control theory.
In the weak system-bath coupling regime where the time-dependent Bloch-Redfield
formalism is applicable, we obtain the Bloch equation to probe the evolution
of the dissipative TLS in the presence
of a time-dependent external control field. By using the automatic differentiation technique
to compute the gradient for the cost functional,
we calculate the optimal transfer integral profile
that can achieve an ideal transfer within a dimer system in the Fenna-Matthews-Olson (FMO) model.
The robustness of the control profile against
temperature variation is also analyzed.
\end{abstract}


\maketitle

\section{Introduction}
The unavoidable coupling to external degrees of freedom and the thereby
caused decoherence still presents the main obstacle for the
realization of a quantum computer~\cite{Nielson,Unruh}.
Several interesting schemes have
been proposed to eliminate
the undesirable effects of decoherence in open quantum systems
such as the use
of decoherence free-subspaces~\cite{Zanardi_1,Lidar},
quantum error correction codes~\cite{Nielson,Preskill,Knill},
quantum Zeno subspace~\cite{Facchi},
quantum dynamical decoupling~\cite{Viola_1,Viola_2,Viola_3,Viola_4,Zanardi_2,Vitali}, and optimized pulse sequence~\cite{Rabitz,NJP,Jirari_1,Jirari_2,Jirari_3,Jirari_4,Jirari_5,Jirari_6}.

The starting point of decoupling techniques
is the observation that even though one does not have access
to the large number of uncontrollable degrees of freedom of the
environment, it is still possible to interfere with its dynamics by
inducing motions into the
system~\cite{Vitali}. This indirect influence of the
environment can be obtained if one can establish an additional coupling
to the system by means of a time-dependent external
control~\cite{Kocharovskaya}. One of the most prominent examples is the application of a sequence of
$\pi$-pulses that flip the sign the qubit-bath coupling operator resulting in the
so-called dynamical decoupling or bang-bang control~\cite{Viola_1}.
The drawback of this scheme is the fact that it eliminates
only noise sources with a frequency below the repetition
of the rate pulses, that is, the decoupling interactions have to be turned
on and off at extremely short time scales, even faster than typical
environment time scales~\cite{Vitali}.

However, these limitations might be circumvented
by using quantum optimal control techniques, which have the great advantage that the decoherence
control can be achieved without moving into the regimes in which
control is much faster than the dynamical time scales of the bath.
Optimal control theory provides a systematic and flexible formalism
that can be used to find the time-optimal pulse sequence
for the manipulation of multi-qubit dissipative systems.

In this work, we apply optimal control theory~\cite{Rabitz,NJP,Jirari_1,Jirari_2,Jirari_3,Jirari_4,Jirari_5,Jirari_6} to the state transfer of a coherently controlled single TLS coupled to an Ohmic bosonic bath through off-diagonal, or bit-flip TLS-bath interactions~\cite{silbey,wu,bitflip}. This system can be described by a driven Ohmic spin-boson model~\cite{hanggi}.

The spin-boson model~\cite{WEISS,Legget} is a widely studied model system
and is relevant to a number of physical situations, including the study of the role
of electron-phonon interaction in point defects and quantum dots,
interacting many-body systems~\cite{Mahan},
magnetic molecules~\cite{Kocharovskaya},
bath assisted cooling of spins and two level Josephson
Junction~\cite{Makhin}, and energy transfer in biological systems~\cite{JPCL,PRE}.
\par Unlike controlling a closed system with unitary dynamical evolution, optimal control of a general open quantum system is a highly nontrivial problem due to the lack of reliable theoretical tools for treating the reduced dynamics of general driven quantum systems under dissipation. However, in the weak or strong system-bath coupling regimes, Redfield-like master equations are believed to be proper methods to deal with driven dissipative dynamics~\cite{PRE,Jirari_3,hanggi,hanggi1}. In Ref.~\cite{Jirari_3}, we applied optimal control theory
to the spin-boson model in the strong system-bath coupling limit where the gradient for the cost functional was obtained by using
the Pontryagin's minimum principle~\cite{Pontryagin} involving a Lagrange multiplier. The application of this approach is possible because the TLS-bath coupling there is diagonal hence the propagator of the coherent system dynamics is trivial. For the off-diagonal TLS-bath coupling, however, if one is still interested in controlling the energy difference of the TLS, then master equations derived in the weak system-bath coupling will generally involve nontrivial propagations of the coherent system. In this case, the application of the Pontryagin's minimum principle to optimal
control problem is less easy. However, thanks to the automatic differentiation method~\cite{tapenade} which allows us to compute the gradient for the cost functional with high precision. Once the gradient is obtained, the minimum of the cost function can be found out by using the conjugate gradient method. The automatic differentiation method has been successfully applied to optimal generation of a single-qubit rotation within the spin-boson model with off-diagonal qubit-bath coupling~\cite{Jirari_2}, where an important property of the cost functional
is its independence of the initial state.
\par By rotating the Hamiltonian around the $y$-axis by an angle of $\pi/2$, the off-diagonal spin-boson model turns out to be capable of describing the dynamics of a strongly coupled dimer system in the Fenna-Matthews-Olson (FMO) protein~\cite{JPCL}. Due to the rotation, the control energy difference profile in the original model is transformed into the transfer integral profile in the FMO model. We perform numerical simulations for the dimer system in the weak exciton-phonon coupling regime and find that our method can achieve an ideal state transfer within physically relevant time scales.

This paper is organized as follows.
In Section \nolinebreak\ref{sec2}
we will review the derivation of Born-Markov master equations for dissipative $N$-level
systems in the presence of time-dependent external control fields.
The master equation is written as a set of Bloch-Redfield equations.
This equation is the starting point for the derivation
of the kinetic equation for the driven spin-boson model in the weak spin--boson coupling
regime, as outlined in Section \nolinebreak\ref{sec3}.
In Section \nolinebreak\ref{sec4} and \nolinebreak\ref{sec5}, we describe the methodology of the optimal control problem and apply the model and method to the study of state transfer in a dimer system in the FMO model. 
Finally, Section \nolinebreak\ref{summary} gives a summary of our findings.
\section{The Bloch-Redfield formalism}\label{sec2}

We begin by reviewing some basic facts about the Bloch-Redfield formalism for general driven dissipative systems in the weak system-coupling limit. Consider a physical system
$S$ embedded in a dissipative environment
$B$
and interacting with a time-dependent classical external control field. The Hilbert space of the total system
${\cal H}_{\rm tot}={\cal H}_S\otimes{\cal H}_B$ is expressed as
the tensor product of the system Hilbert space ${\cal H}_S$
and the environment Hilbert space ${\cal H}_B$.
Here, we suppose that ${\cal H}_S$ is $N$-dimensional with some time-independent
orthonormal basis $\{|i\rangle\}, i=1,2\ldots N$.
The total Hamiltonian has the general form
\begin{eqnarray}
H_{\rm tot} =H_c(t)+ H_B+ H_{\rm int},
\label{Hamiltonian1}
\end{eqnarray}
where $H_c(t)$ is the system part of the Hamiltonian, $H_B$
describes the bath, and $H_{\rm int}$ is the system-bath interaction
that is responsible for decoherence.
The operators $H_c(t)$ and $H_B$ act on ${\cal H}_c$ and  ${\cal H}_B$, respectively. The system Hamiltonian $H_c(t)$ is explicitly time-dependent through the external control field. The system-environment interaction is assumed to be of bilinear form
$
H_{\rm int}= \sum_{\alpha}\, A_\alpha\otimes B_\alpha
$

with $A_\alpha$ and $B_\alpha$ Hermitian operators of the system and the environment, respectively.

In order to investigate decoherence in the limit of weak system-bath coupling,
the Bloch-Redfield formalism can be used to derive a set of a master
equations for the reduced density matrix
$\rho_S(t)={\rm tr}_{\rm{B}}\left\{\rho_{\rm tot}(t)\right\}$
describing the system dynamics, where
$\rho_{\rm tot}$ is the total density matrix for both the system
and the bath. Starting from the Liouville-von Neumann equation
$
i\dot\rho_{\rm tot}(t)=\left\lbrack H_{\rm tot},\rho_{\rm tot}(t)\right\rbrack
$
for the total density operator and after performing Born and Markov
approximations, one obtains the Bloch-Redfield master equation for $\rho_S(t)$
in the basis $\{|i\rangle\}$~\cite{Jirari_3,Kocharovskaya,hanggi}
\begin{eqnarray}
\label{eq:MASTER_EQUATION}
{\dot\rho}_{S,ij}(t)&=-&\frac{i}{\hbar}\sum_{kl}\left(H_{S,ik}(t)\delta_{lj}
-\delta_{ik}H_{S,lj}(t)\right){\rho}_{S,kl}(t)
\nonumber\\
&&-\sum_{kl}{\cal R}_{ijkl}(t){\rho}_{S,kl}(t),
\end{eqnarray}
where the first term on the right hand side represents the unitary part of the dynamics
generated by the system Hamiltonian $H_S(t)$ and the second term
accounts for dissipative effects of the coupling to the environment.
The Redfield relaxation tensor $R_{ijkl}(t)$ is given by
\begin{eqnarray}
\label{eq:REDFIELD_TENSOR}
{\cal R}_{ijkl}(t)&=&\delta_{lj}\sum_r\,\Gamma_{irrk}^+(t)
+\delta_{ik}\sum_r\,\Gamma_{lrrj}^-(t)
\nonumber\\
&&-\Gamma_{ljik}^+(t) -\Gamma_{ljik}^-(t),
\end{eqnarray}
where the time-dependent rates $\Gamma_{ijkl}^{\pm}(t)$ are evaluated
as
\begin{eqnarray}
\label{eq:Gamma+_BIS}
&&\Gamma_{lj,ik}^+(t)=\frac{1}{\hbar^2}\int_0^{t} dt'
\sum_{\alpha,\beta}
\left\langle B_\alpha(t-t')B_\beta(0)\right\rangle_B
\nonumber\\
&&A_{\alpha,{lj}}\times\sum_{m,n}
U^c_{{im}}(t,t')A_{\beta,{mn}}U^{c\ast}_{{kn}}(t,t'),\nonumber\\
&&\Gamma_{lj,ik}^-(t)=\frac{1}{\hbar^2}\int_0^{t} dt'
\sum_{\alpha,\beta}
\left\langle B_\beta(0)B_\alpha(t-t')\right\rangle_B
\nonumber\\
&&\times
\sum_{m,n} U^c_{{lm}}(t,t')A_{\beta,{mn}}U^{c\ast}_{{jn}}(t,t')\,A_{\alpha,{ik}},
\end{eqnarray}
with
\begin{eqnarray}
\label{eq:PROPAGATOR_SYS}
U^c(t,t') = {\cal T}\left\{\exp\left\lbrack -{i\over\hbar}\int_{t'}^t
\,d\tau\,H_c(\tau)\right\rbrack\right\}
\end{eqnarray}
being the propagator of the coherent system dynamics satisfying the Schr\"odinger equation
\begin{eqnarray}
\label{eq:SCHRODINGER_EQUATION}
i\hbar\frac{\partial}{\partial t} U^c(t,t')= H_c(t)U^c(t,t'),
~ U^c(t',t')= {\cal I}.
\end{eqnarray}
In Eqs. (\ref{eq:Gamma+_BIS}), the environment correlation functions read
\begin{eqnarray}
\langle B_\alpha(\tau)B_\beta(0)\rangle_B ={\rm tr}_{\rm{B}}\left\{B_\alpha(\tau)B_\beta(0)\rho_B\right\},
\end{eqnarray}
where $\rho_{\rm{B}} = \exp(-\beta H_B)/Z_B$ is the thermal equilibrium density matrix of the bath with the inverse temperature $\beta=1/k_BT$ and the partition function $Z_B=\rm{tr}_B\{\rho_B\}$. Eq.~(\ref{eq:MASTER_EQUATION}) was obtained under the
assumption that
$
\left\langle B_\alpha(\tau)\right\rangle_B={\rm tr}_{\rm{B}}\left\{ B_\alpha(\tau)\rho_B\right\}=0,
$
which states that the reservoir averages of $B_\alpha(\tau)$ vanish. Note that the time-dependent control field which enters $H_c(t)$ is treated
non-perturbatively in the derivation of the master equation. Note also that the time-dependent control field enters the dissipative part of the evolution as well through the field-dependent relaxation rates $\Gamma_{ijkl}^\pm(t)$ via
$U^c(t,t')$, which is a consequence of quantum interference between the system-bath coupling and the external coupling to the control field. This allows for an external
control of dissipation~\cite{Jirari_3,Kocharovskaya}.
In the next section, we will apply the foregoing
formalism to a TLS coupled to a boson bath
via weak off-diagonal TLS-bath coupling, and
derive the corresponding Bloch equations
satisfied by the Bloch vector of the TLS.
\section{Model and master equation}\label{sec3}

The driven spin-boson model in which a
qubit or a spin-$1/2$ coupled to a bosonic bath via off-diagonal spin-bath coupling
and subjected to a time-dependent external force
can be written as
\begin{eqnarray}
H_\mathrm{tot}(t)&=&H_\mathrm{s}+H_\mathrm{c}(t)+H_\mathrm{b}+H_\mathrm{sb},\nonumber\\
H_\mathrm{s}&=&-\frac{\varepsilon_0}{2}\sigma_z-\frac{\Delta}{2}\sigma_x,~H_\mathrm{c}(t)=-\frac{\varepsilon(t)}{2}\sigma_z,\\
H_\mathrm{b}&=&\sum_{i}\omega_ib_i^\dag b_i,~H_\mathrm{sb}=\frac{\sigma_x}{2}\sum_{i}  c_i(b_i^\dag+b_i),\nonumber
\end{eqnarray}

Here, $H_\mathrm{s}$ is the bare TLS Hamiltonian with energy level spacing $\varepsilon_0$ (with $\sigma_{x,z}$ being the Pauli matrices in the computational basis $\{|0\rangle,|1\rangle\}$ of the TLS), $\Delta$ measures the transfer integral between the two levels. $H_\mathrm{c}(t)$ specifies the external control term with varying energy difference $\varepsilon(t)$. $H_\mathrm{b}$ describes the free boson bath with annihilation operator $b_i$ of the bath mode with frequency $\omega_i$.
$H_\mathrm{sb}$ captures the off-diagonal coupling of the bath with the TLS with strength $c_i$.

In order to investigate the TLS dynamics
in the limit of weak system-bath coupling,
the Bloch-Redfield formalism developed in the last section
can be readily used to derive a set of a master
equations for the qubit density matrix
$\rho_{\rm s}(t)={\rm tr}_{\rm B}\left\{\rho_{\rm tot}(t)\right\}$.
For the two-level system studied here,
it is convenient to study the dynamics of the Bloch vector
$\mathbf{p}(t)=\left(p_x(t),p_y(t),p_z(t)\right)^T\in {\mathbb{R}^3}$
defined by\break
$
\mathbf{p}(t)= {\rm tr}_{\rm S}(\vec{\mbox{$\sigma$}}\rho_{\rm s}(t)).
$

Straightforward application of Eq. (\ref{eq:MASTER_EQUATION}) to the TLS in its computational basis gives the following set of Bloch equations,
\begin{eqnarray}\label{Bloch}
\dot{\mathbf{p}}(t)=M(t)\mathbf{p}(t)+R(t),
\end{eqnarray}
with
\begin{eqnarray}\label{M_R}
M(t)&=&\left(
       \begin{array}{ccc}
         0 & \varepsilon_0+\varepsilon(t) & 0 \\
         -[\varepsilon_0+\varepsilon(t)+\Gamma_{yx}(t)] & -\Gamma_{yy} & \Delta \\
         -\Gamma_{zx}(t) & -\Delta & -\Gamma_{zz}(t) \\
       \end{array}
     \right),\nonumber\\
R(t)&=&(0,-A_y(t),-A_z(t))^T.
\end{eqnarray}
Following the notation used in~\cite{hanggi},
the fluctuating terms in the inhomogeneous part $R(t)$ are given by
\begin{eqnarray}\label{Ayz}
A_y(t)&=&2\int_0^t dt'{{\mathcal{M}}}''(t-t')
\mbox{Re}\left\lbrack U_{{11}}(t,t')U^{*}_{{12}}(t,t')\right\rbrack,\nonumber\\
A_z(t)&=&\int_0^t dt'{{\mathcal{M}}}''(t-t')
\mbox{Re}\left\lbrack U_{{11}}(t,t')^2-U_{{12}}(t,t')^2\right\rbrack,\nonumber\\
\end{eqnarray}
and the temperature dependent relaxation rates are determined
by
\begin{eqnarray}\label{gammaij}
\Gamma_{ij}(t)=\int_0^t dt'{{\mathcal{M}}}'(t-t')b_{ij}(t,t'),
\end{eqnarray}
with $\Gamma_{zz}(t)=\Gamma_{yy}(t)$.

In Eqs.~(\ref{Ayz}) and (\ref{gammaij}), the functions ${{\mathcal{M}}}'$ and ${{\mathcal{M}}}''$
are the real part and imaginary part, respectively, of the bath correlation function
\begin{eqnarray}\label{Mt}
M\left(t\right)=\frac{1}{\pi}\int_{0}^{\infty}d\omega
J(\omega)\frac{\cosh(\frac{\beta\omega}{2}-i\omega
t)}{\sinh(\frac{\beta\omega}{2})},
\end{eqnarray}
The functions $b_{ij}(t,t')$ read
\begin{eqnarray}\label{bij}
b_{yx}(t,t')&=&\mbox{Im}\left\lbrack U_{{11}}^2(t,t')-U_{{12}}^2(t,t')\right\rbrack,\nonumber\\
b_{yy}(t,t')&=&\mbox{Re}\left\lbrack U_{{11}}^2(t,t')-U_{{12}}^2(t,t')\right\rbrack,\nonumber\\
b_{zx}(t,t')&=&-2\mbox{Re}\left\lbrack U_{{11}}(t,t')U^{*}_{{12}}(t,t')\right\rbrack,
\end{eqnarray}
where $U(t,t')$ is the non-dissipative time evolution operator for the spin system
with $U_{{11}}(t,t')=\langle 0|U(t,t')|0\rangle$
and $U_{{12}}(t,t')=\langle 0|U(t,t')|1\rangle$, and satisfies the Schr\"odinger equation
\begin{eqnarray}
\label{eq:UEM}
\dot U(t,0)
=\frac{i}{2}\left\lbrack\Delta{\sigma_x} +\left(\varepsilon_0+\varepsilon(t)\right){\sigma_z}
\right\rbrack\,{U}(t,0).
\end{eqnarray}

In this work, we will employ the Ohmic spectral density
$J(\omega)=2\pi\alpha\omega e^{-\omega/\omega_c}$ for the bath,
where $\alpha$ is a dimensionless coupling constant
and $\omega_c$ is the bath's cutoff frequency.
In the next section, we will apply quantum optimal
control theory to the Bloch equations Eq.(\ref{Bloch})
and determine the optimal energy difference profile of the TLS that can
give a perfect population inversion of the TLS within
certain time interval.
\section{Quantum optimal control problem}\label{sec4}
Our aim here is to present a general method that allows us to solve the
following inverse problem: what is the time-dependent control field under which
the Bloch vector can switch from a given initial state to a given prescribed target
state within a fixed time interval $t\in[0,t_F]$? Now we formulate this problem in the framework of optimal control theory~
\cite{NJP,Jirari_1,Jirari_2,Jirari_3,Jirari_4,Jirari_5}.
Suppose the system is prepared at time $t_I=0$ in the initial state
$\mathbf{p}(0)=\mathbf{p}_I$. The objective is to compute
an appropriate time-dependent control function ${\varepsilon}(t)$
steering the system from the initial state $\mathbf{p}_I$
into a desired state $\mathbf{p}_D$ at time $t_F$.
This goal leads to the following optimal control problem:
determine a continuously differentiable function
${\varepsilon}(t),~t\in[0,t_F]$, which gives the minimal value
of the following cost functional
\begin{eqnarray}
J(\varepsilon)= \frac{1}{2}\Vert\mathbf{p}(t_F)-\mathbf{p}_D\Vert_2^2+
\frac{\nu}{2}\int_0^{t_F}\,dt \varepsilon^2(t),
\end{eqnarray}
while, at the same time, satisfies the dynamic constraints
and the boundary conditions Eqs.~(\ref{Bloch})-(\ref{eq:UEM}).

Here, the cost functional $J(\varepsilon)$ represents the deviation
of the state of the system at final time $\mathbf{p}(t_F)$ from
the desired state $\mathbf{p}_D$. Minimizing $J(\varepsilon)$ leads to
the physical target we want to reach.
The second integral penalizes the field fluency
$\mathcal{E}=\int_0^{t_F}\,dt~\varepsilon^2(t)$ with weight $\nu > 0$.

In principle one can use
the Pontryagin's minimum principle
to treat our optimal control problem and derive the gradient for the cost
functional $J(\varepsilon)$~\cite{NJP,Jirari_1,Jirari_2,Jirari_3,Jirari_4,Jirari_5}.
However, for the off-diagonal spin-boson model studied here, the response of the system to the variation of the control
$\varepsilon(t)$ is determined by the master equation Eq.~(\ref{Bloch})
and the equation of motion for the propagator of the coherent system dynamics
Eq.~(\ref{eq:UEM}). As a result, the application of
the Pontryagin's minimum principle
is less straightforward since
two Lagrange multipliers have to be
introduced to implement these two dynamical constraints.

An alternative to this approach is the technique
of automatic differentiation~\cite{tapenade} which in principle amounts to doing
calculus on the fully discretized form of the optimal control problem.
For this purpose, we firstly
discretize the time interval $I=[0,t_F]$
into $M$ equal-sized subintervals $\Delta I_k$
with $I=\bigcup_{k=1}^M \Delta I_k$ and then approximate $\varepsilon(t)$ as
$\varepsilon(t) \rightarrow \varepsilon(t_k)=\varepsilon_k,\, k=1\ldots M$.
Thus the problem becomes that of finding
$\vec\varepsilon=\left(\varepsilon_1,
\ldots,\varepsilon_M\right)^T\in\mathbb{R}^M$
such that $J(\vec\varepsilon)=\mbox{inf}\left\{J(\vec\zeta): \vec\zeta\in\mathbb{R}^M\right\}$.
Automatic differentiation tools can be viewed as black boxes taking
as input a program computing the cost function $J(\vec\varepsilon):
\mathbb{R}^M\longrightarrow\mathbb{R}$
and giving as output another program computing
the gradient $\partial J/\partial\vec\varepsilon\in\mathbb{R}^M$.
Two approaches to automatic differentiation are possibles : the forward (or tangent) mode
and backward (or adjoint) mode which is similar to the adjoint method~\cite{tapenade}. In this work, we employ the latter since it is theoretically more efficient
in computing the gradient of a scalar value function.
With the gradient obtained from the adjoint mode
of automatic differentiation, the optimization of
the cost function $J(\vec\varepsilon)$
is then performed by using the conjugate
gradient algorithm.

Before starting the discussion of our numerical results,
let us first analyze the master equation without the control field, $\varepsilon(t)=0$. For this undriven case,
the analytical expression of the matrix elements for the coherent propagator are given by
~\cite{Igor}

\begin{eqnarray}
U_{{11}}(\tau)&=&\cos(\Omega\tau/2)+i\epsilon_0\sin(\Omega\tau/2)/\Omega,\nonumber\\
U_{{12}}(\tau)&=&i\Delta\sin(\Omega\tau/2)/\Omega,
\end{eqnarray}
where $\Omega=\sqrt{\epsilon_0^2+\Delta^2}$.
The decay rates follow as
\begin{align}
\label{eq:gammayy}
&\Gamma_{yy}=\frac{\Delta^2}{\Omega^2}S(0)
+\frac{\epsilon_0^2}{2\Omega^2}S(\Omega),\quad
\Gamma_{yx}=-2\alpha\epsilon_0\log(\omega/\Omega)
\nonumber
\\
&\Gamma_{zx}=\frac{\Delta\epsilon_0}{\Omega^2}\left(S(\Omega)/2-S(0)\right),\qquad
A_z = -\epsilon_0\pi\alpha,
\\
\label{eq:ay}
&A_y = -\frac{2\alpha\omega\Delta\epsilon_0}{\Omega^2}.
\nonumber
\end{align}
with $S(\omega)=J(\omega)\coth(\beta\omega/2)$.
The stationary state solution of the master equation in the long-time limit $t\to\infty$ can be readily obtained as
\begin{eqnarray}
\label{eq:sstate}
\mathbf{p}^T(\infty)=
\left(
{\Delta}/{\Omega},
0,
{\varepsilon_0}/{\Omega}
\right)\times\tanh(\beta\Omega/2).
\end{eqnarray}
\section{Application to state transfer in light-harvesting systems}\label{sec5}
As an illustration, we now apply our approach to the optimal energy transfer in a widely studied photosynthetic system, the Fenna-Matthews-Olson (FMO) protein~\cite{FMO1,FMO2}. The complex is a trimer formed by three monomers, each of which contains seven bacteriochlorophyll-$a$ (BChla) chromophores. It is known that the transfer integral between the BChl$a$ 1 and BChl$a$ 2 is the strongest among other electronic coupling strengths~\cite{JPCL,renger}. Thus, in the following, we will focus on the energy transfer in this dimer system composed by these two chromophores. In the single-excitation subspace, the dimer is described by the following rotated Hamiltonian~\cite{JPCL}
\begin{eqnarray}
\tilde{H}&=&e^{i\pi\sigma_y/4}He^{-i\pi\sigma_y/4}\nonumber\\
&=&\frac{\varepsilon_0+\varepsilon(t)}{2}\sigma_x-\frac{\Delta}{2}\sigma_z+\frac{\sigma_z}{2}\sum_ic_i(b_i+b^\dag_i)+H_b,\nonumber\\
\end{eqnarray}
where the control field now becomes the fluctuating transfer integral on top of the bare one $\varepsilon_0$.

We now consider the task of driving the system from the population state $|i\rangle=|1\rangle$
into a desired superposition state
$|f\rangle=\frac{1}{\sqrt{2}}\left(|1\rangle+i|0\rangle\right)$. Such a coherent state has been considered before in the study of electronic coherence in the FMO complexes~\cite{nphy2013}. The above state transfer is equivalent to the population inversion in the unrotated frame, i.e., a transfer from the initial state $\mathbf{p}_I=(1,0,0)^T$ into the desired state
$\mathbf{p}_D :=(0,1,0)^T$ at time $t_F$, since
$e^{i\pi\sigma_y/4}|1_x\rangle=|1\rangle$ and
$e^{i\pi\sigma_y/4}|1_y\rangle=
\frac{1}{\sqrt{2}}(e^{-i\pi/4}|1\rangle+e^{i\pi/4}|0\rangle)$. In the
numerical simulations, we employ the following parameters for the FMO
complex~\cite{JPCL}: the energy gap $\Delta=-75 cm^{-1}$, the bare
transfer integral $\varepsilon_0=175.4 cm^{-1}$, $1/\beta=53.5cm^{-1}$
(or $T=77K$), $\omega_c=166.7 cm^{-1}$, $\alpha=10^{-3}$, and
$t_F=0.094/(cm^{-1})$ (or $t_F=500 fs$). We stress that in order to
ensure the validity of the Bloch-Redfield formalism, we have set the
dimensional exciton-phonon coupling strength $\alpha$ to be much
smaller than its realistic value ($\alpha\approx0.1$) in the FMO
model. The numerical simulations are thus only for illustration
purpose.
\begin{figure}[htbp]
\includegraphics[width=7cm,angle=-90]{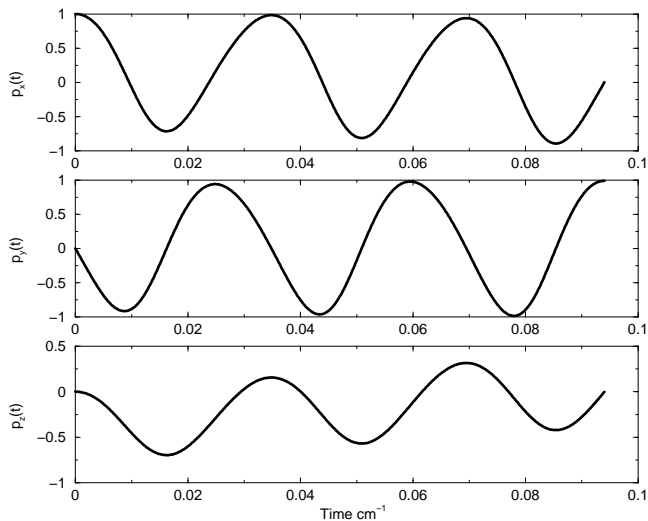}
\caption{(Color online)
Transfer from the initial state
$\vec{\bf p}_I=(1,0,0)^T$ into target state $\vec{\bf p}_D=(0,1,0)^T$ in the unrotated frame, which corresponds to the transfer from the initial state
$|i\rangle$ into target state $|f\rangle$
under the optimal transfer integral profile of the dimer system.
Parameters (in unit of $cm^{-1}$):
$\varepsilon_0=175.4,~\Delta =-75$, $\omega_c=166.7$,
$\frac{1}{\beta}=53.5$, $t_F=0.094$.
Dimensionless parameters : $\alpha=10^{-3}$ and $\nu=10^{-3}$.
}
\label{fig:fig1}
\end{figure}
\begin{figure}[htbp]
\includegraphics[width=7cm,angle=-90]{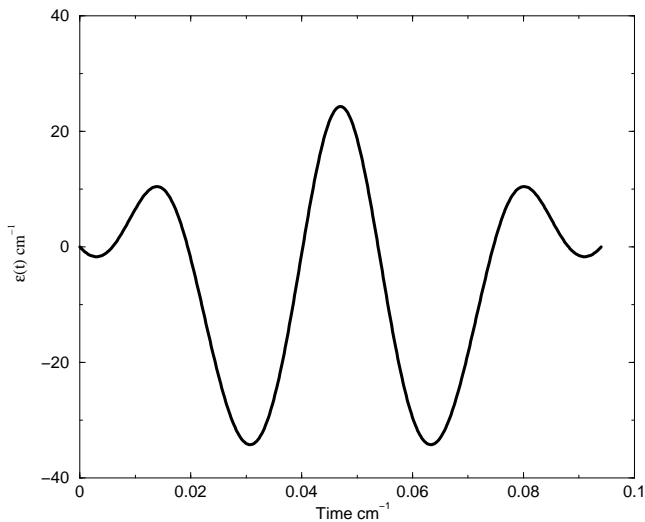}
\caption{(Color online)
The optimized control field $\varepsilon_{\rm{opt}}(t)$
for achieving a perfect state transfer from the initial state
$|i\rangle=|1\rangle$
to the desired superposition state
$|f\rangle=\frac{1}{\sqrt{2}}\left(|1\rangle+i|0\rangle\right)$ in the dimer system.
The parameters are the same as that in Fig.~\ref{fig:fig1}.}
\label{fig:fig2}
\end{figure}


The time evolution of the components of the Bloch vector (in the unrotated frame)
under the influence of the optimized control field is shown
in Fig.~\ref{fig:fig1}. We see that perfect state transfer is achieved.
The corresponding optimized control field $\varepsilon_{\rm{opt}}(t)$ is shown in Fig.~\ref{fig:fig2}.
We note that 
the optimal profile varies smoothly over the time interval considered, which makes coherent control of the state transfer possible in practical situations.

\begin{figure}[htbp]
\includegraphics[width=6cm,angle=-90]{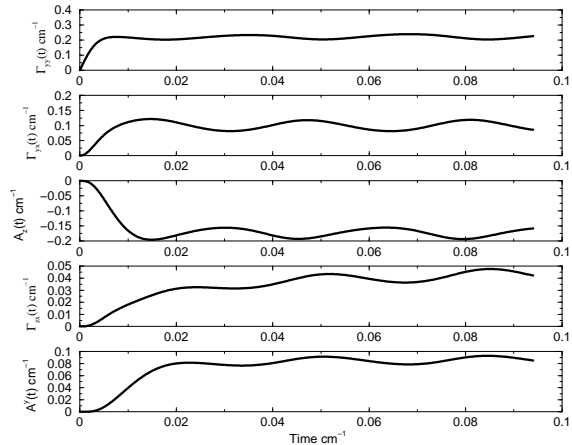}
\caption{(Color online) The time dependence of the decay rates
under the influence of the optimized control field.
The parameters are the same as that in Fig.~\ref{fig:fig1}.}
\label{fig:fig3}
\end{figure}

In Fig.~\ref{fig:fig3} we plot the time evolutions of the decay rates
appearing in the Bloch equation, under the optimal control field
$\varepsilon_{\rm{opt}}(t)$. In contrast to the undriven case where
the rates are given by there stationary values Eq.(\ref{eq:gammayy}),
the dynamics of these controlled rates reflects the temporal structure of the optimal control.

In practical cases, the system parameters are usually not fixed but subject
to external perturbations or noises. Because of these factors, practical devices are generally not
capable of operating precisely at the computed control field.
So it is of great importance to know the sensitivity of the optimal solution with
respect to perturbations of the system parameters. The spin-boson
Hamiltonian depends on many parameters, namely the bias energy
$\varepsilon_0$, the tunneling splitting $\Delta$, the coupling strength
$\alpha$, the bath's cutoff frequency $\omega_c$ and the temperature $1/\beta$ , etc.
For demonstration purposes, here we consider the effect of slow fluctuations
of the bath temperature $1/\beta$ under the optimal control pulse.
We apply the optimal control to an ensemble of systems
with normal variation in the parameter $\beta$~\cite{Grace}
and analyses how the uncertainties in the system parameter $\beta$ affect
the quantum purity $\mathcal{P}=\|\vec{\bf p}(t_F)\|^2$.

Our statistical analysis employs only mean values and standard deviations
given by $\overline{\mathcal{P}}=\frac{1}{i_0}\Sigma_{i=1}^{i_0}\mathcal{P}_i$
and $\sigma_\mathcal{P}=\lbrack\frac{1}{i_0}\Sigma_{i=1}^{i_0}
({\mathcal{P}}_{i}-\overline{\mathcal{P}})^2\rbrack^{1/2}$.
The value of each parameter $\beta$
is individually replaced by a value randomly selected from a
normal distribution
with a mean ${\overline\beta}= (53.5 cm^{-1})^{-1}$
and a standard deviation $\sigma_\beta=\frac{\overline\beta}{10}$.
We find the following average purity
$\overline{\mathcal{P}}=0.98605136$
and the standard deviation
$\sigma_\mathcal{P}= 0.00948239$.
The high value of the average purity
indicate that the optimal pulse we obtained is indeed
robust with random temperature variations.

\section{Summary}\label{summary}

In this work, we considered the optimal
state transfer problem of a single two-level
system coupled to a boson bath through off-diagonal
TLS-bath coupling. In the weak system-bath coupling limit,
we use the Bloch equation for the Bloch vector of the
TLS under general time-dependent coherent control field
acting on the TLS. We demonstrate the importance
of automatic differentiation method in evaluating
the gradient of the cost functional, because of the
nontrivial coherent evolution of the TLS.
We then apply our method to study the coherent state transfer in a
dimer system within the FMO complex.
Perfect transfer from an initial occupation state into a superposition state is achieved in physical time scales.
The present method can also be applied to the study
of state preparations in multi-qubit dissipative systems.

\noindent{\bf Author contribution statement}\\
H.J. is the main contributor of the work. Both of the authors contributed to the writing of the paper.


\begin{thebibliography}{99}
\bibitem{Nielson}
M. A. Nielson and I. L. Chuang, Quantum Computation and Quantum Information
(Cambridge University Press, Cambridge, UK, 2000).
\bibitem{Unruh}
 W. G. Unruh, Phys. Rev. A {\bf 51}, 992 (1995).
\bibitem{Zanardi_1}
 P. Zanardi and M. Rasetti, Phys. Rev. Lett. {\bf 79}, 3306 (1997).
\bibitem{Lidar}
D. A. Lidar, I. L. Chuang and K. B. Whaley,
Phys. Rev. Lett. {\bf 81}, 2594 (1999).
\bibitem{Preskill}
J. Preskill,  Proc. Roy. Soc. Lond. A{\bf 454}, 385 (1998).
\bibitem{Knill}
E. Knill E, R. Laflamme and L. Viola,
Phys. Rev. Lett. {\bf 84}, 2525 (2000).
\bibitem{Facchi}
P. Facchi and Pascazio, Phys. Rev. A {\bf 89}, 080401 (2002).
\bibitem{Viola_1}
L. Viola and S. Lloyd, Phys. Rev. A {\bf 58}, 2733 (1998).
\bibitem{Viola_2}
L. Viola, E. Knill, and S. Lloyd, Phys. Rev. Lett. {\bf 82}, 2417 (1999).
\bibitem{Viola_3}
L. Viola, E. Knill, and S. Lloyd, Phys. Rev. Lett. {\bf 83}, 4888 (1999).
\bibitem{Viola_4}
L. Viola, E. Knill, and S. Lloyd, Phys. Rev. Lett. {\bf 85}, 3520 (2000).
\bibitem{Zanardi_2}
P. Zanardi, Phys. Lett. A 258, 77 (1999).
\bibitem{Vitali}
D. Vitali and P. Tombesi, Phys. Rev. A {\bf 59}, 4178 (1999).
\bibitem{Rabitz}
I. Walmsley and H. Rabitz, Physics Today {\bf 56}, 43-49 (2003).
\bibitem{NJP}
 C. Brif, R. Chakrabarti, and H. Rabitz, New J. Phys. \textbf{12}, 075008 (2010).
\bibitem{Jirari_1}
H. Jirari and W. P\"otz, Phys. Rev. A {\bf 72}, 013409 (2005).
\bibitem{Jirari_2}
H. Jirari, Euro. Phys. Lett. {\bf 87 } 40003 (2009).
\bibitem{Jirari_3}
H. Jirari and W. P\"otz, Phys. Rev. A {\bf 74}  022306 (2006).
\bibitem{Jirari_4}
 H. Jirari, F. W. J. Hekking and O. Buisson, Euro. Phys. Lett. {\bf 87 } 28004 (2009).
\bibitem{Jirari_5}
H. Jirari and W. P\"otz, Euro. Phys. Lett. {\bf 77 }  40003 (2007).
\bibitem{Jirari_6}
N. Barros, M. Rassam, H. Jirari and H. Kachkachi,
Phys. Rev. B {\bf 83},  144418 (2011).
\bibitem{Kocharovskaya}
O. Kocharovskaya, S.-Y Zhu, M. O. Scully, P. Mandel, and Y. V. Radeonychev, Phys. Rev. A {\bf 49}, 4928 (1994).
\bibitem{silbey}
R. W. Munn and R. Silbey, J. Chem. Phys. \textbf{83}, 1843 (1985).
\bibitem{wu}
N. Wu, K.-W Sun, Z. Chang, and Y. Zhao, J. Chem. Phys. \textbf{136}, 124513 (2012).
\bibitem{bitflip}
K. M. Fonseca-Romero, S. Kohler, and P. H\"anggi, Phys. Rev. Lett. {\bf 95}, 140502 (2005).
\bibitem{hanggi}
L. Hartmann, I. Goychuk, M. Grifoni, and P. H\"anggi, Phys. Rev. E \textbf{61}, R4687(R), (2002).
\bibitem{WEISS}
U. Weiss, Quantum Dissipative Systems, (World Scientific, Singapore 1999).
\bibitem{Legget}
A. J. Leggett, S. Chakravarty, A. T. Dorsey, M. P. A. Fisher,
A. Garg and W. Zwerger, Rev. Mod. Phy {\bf 59}, 1 (1987).
\bibitem{Mahan}
G. D. Mahan, Many-Particle Physics, 3rd ed. (Kluwer Academic, Dordrecht 2000).
\bibitem{Makhin}
Y. Maklin, G. Sch\"on and and  A. Schnirman Mod. Phy {\bf 73}, 357 (2001).
\bibitem{JPCL}
L. A. Pach\'on and P. Brumer, J. Phys. Chem. Lett. \textbf{2}, 2728 (2011).
\bibitem{PRE}
J. Cai, S. Popescu, and H. J. Briegel, Phys. Rev. E \textbf{82}, 021921 (2010).
\bibitem{hanggi1}
D. Zuecok, P. H\"anggi, and S. Kohler, New J. Phys., \textbf{10}, 115012 (2008).
\bibitem{Pontryagin}
L. S. Pontryagin, {\it The Mathematical Theory of the Optimal Process}
(Wiley-Interscience, New York, 1962).
\bibitem{tapenade}
L. Hascoet and V. Pascual, TAPENADE 2.1 user's guide,
Rapport technique 0300, INRIA, September (2004).
\bibitem{Igor}
I. Goychuk, P. H\"anggi, Advances in Physics, {\bf 54}, 525 (2005).
\bibitem{FMO1} R. E. Fenna, B. W. Matthews, Nature, \textbf{258}, 573 (1975).
\bibitem{FMO2} A. Camara-Artigas, R. E. Blankenship, J. P. Allen, Photosynth. Res. \textbf{75}, 49 (2003).
\bibitem{renger} J. Adolphs and T. Renger, Biophys. J. \textbf{91}, 2778 (2006).
\bibitem{nphy2013} A. W. Chin, \emph{et. al.}, Nat. Phys. \textbf{9}, 113 (2013).
\bibitem{Grace}
M. Grace, C. Brif, H. Rabitz, I. A. Walmsley, R. L. Kosut, D. A. Lidar, J. Phys. B: At. Mol. Opt. Phys. \textbf{40}, S103 (2007).
\end{thebibliography}
\end{document}